\documentclass[letterpaper,12pt]{article}
\usepackage{array}
\usepackage{dcolumn}
\usepackage{float}
\usepackage[numbers]{natbib}
\usepackage{graphicx}
\usepackage{color}
\usepackage{amsmath}
\usepackage{amsfonts}
\usepackage{amsopn}
\usepackage{amsthm}
\usepackage{lscape}
\usepackage{multirow}
\newtheorem{thm}{Theorem}

\newtheorem{prop}[thm]{Proposition}
\theoremstyle{definition}

\theoremstyle{remark}

\DeclareMathOperator{\pr}{P}
\DeclareMathOperator{\epn}{E}
\DeclareMathOperator{\var}{var}
\DeclareMathOperator{\avar}{avar}

\DeclareMathOperator{\expit}{expit}

\newcommand{\bmX}{\boldsymbol X}
\newcommand{\bmO}{\boldsymbol O}

\newcommand{\bmr}{\boldsymbol r}
\newcommand{\bmb}{\boldsymbol b}
\newcommand{\bmc}{\boldsymbol c}
\newcommand{\bmd}{\boldsymbol d}
\newcommand{\bmzero}{\boldsymbol 0}
\newcommand{\bmone}{\boldsymbol 1}
\newcommand{\bmalpha}{\mbox{\boldmath${\alpha}$}}
\newcommand{\bmbeta}{\mbox{\boldmath${\beta}$}}
\newcommand{\bmgamma}{\mbox{\boldmath${\gamma}$}}

\newcommand{\bmpsi}{\mbox{\boldmath${\psi}$}}
\newcommand{\bmphi}{\mbox{\boldmath${\phi}$}}
\newcommand{\bmnu}{\mbox{\boldmath${\nu}$}}
\newcommand{\mbI}{\mathbf I}

\renewcommand{\baselinestretch}{1}

\begin{document}
\begin{center}
{\LARGE\bf A Model-Robust G-Computation Method for Analyzing Hybrid Control Studies Without Assuming Exchangeability}
\end{center}

\vspace{0.35cm}
\begin{center}
Zhiwei Zhang$^{1,*}$, Peisong Han$^1$ and Wei Zhang$^2$\\
$^1$Biostatistics Innovation Group, Gilead Sciences, Foster City, California, USA\\
$^2$State Key Laboratory of Mathematical Sciences, Academy of Mathematics and Systems Science, Chinese Academy of Sciences, Beijing, China\\
$^*$zhiwei.zhang6@gilead.com
\end{center}

\vspace{0.35cm}
\centerline{\bf Abstract}

There is growing interest in a hybrid control design for treatment evaluation, where a randomized controlled trial is augmented with external control data from a previous trial or a real world data source. The hybrid control design has the potential to improve efficiency but also carries the risk of introducing bias. The potential bias in a hybrid control study can be mitigated by adjusting for baseline covariates that are related to the control outcome. Existing methods that serve this purpose commonly assume that the internal and external control outcomes are exchangeable upon conditioning on a set of measured covariates. Possible violations of the exchangeability assumption can be addressed using a g-computation method with variable selection under a correctly specified outcome regression model. In this article, we point out that a particular version of this g-computation method is protected against misspecification of the outcome regression model. This observation leads to a model-robust g-computation method that is remarkably simple and easy to implement, consistent and asymptotically normal under minimal assumptions, and able to improve efficiency by exploiting similarities between the internal and external control groups. The method is evaluated in a simulation study and illustrated using real data from HIV treatment trials.

\vspace{.5cm}

\section{Introduction}\label{sec:intro}

Randomized controlled trials (RCTs) are the gold standard for evaluating treatment safety and effectiveness, as randomization balances both observed and unobserved baseline covariates and supports unbiased estimation of treatment effects. However, in settings such as rare diseases, relying solely on randomized data may be inefficient or infeasible \citep{m03,j18}. These challenges have motivated the use of external control data from prior studies or real‑world sources \citep{fda23a}. The hybrid control design, which augments an RCT with external control data, can improve precision and power, reduce cost, and potentially facilitate enrollment in the RCT. Yet, systematic differences between external and randomized populations may introduce bias and inflate type I error if not properly addressed.

A variety of methods have been proposed to address potential discrepancies between internal and external control groups. A common strategy is to discount the external control group before combining it with the internal control group. Discounting is usually conducted in a Bayesian framework, such as through power priors \citep{ic00,n10}, but also can be done in a frequentist \citep{t22,l23} or hybrid \citep{w19,c20,lu22,f23,w23} manner. The discounting factor (e.g., the power parameter in a power prior) can be determined adaptively using Bayesian hierarchical models \citep{h11,v14}, empirical Bayes approaches \citep{gh17}, or frequentist techniques \citep{t22}. Roughly speaking, adaptive discounting tailors the contribution of external controls to the observed level of agreement with the internal controls, applying heavier discounting when the groups differ more. The theoretical properties of discounting methods, such as consistency and efficiency, are not well established in the current literature.

Another general approach to the hybrid control design is to adjust for prognostic baseline covariates that may drive differences between internal and external control groups. These methods draw heavily from the causal inference literature \citep{r86,rr84,rr85,r00,vr03,vr11}. Some of these methods rely on a propensity score (PS) model \citep{rr83}, which in this context may be defined as the conditional probability (based on covariate values) that a control subject in the study originates from the
RCT. Estimated PS values can be used for matching, stratification, or weighting. Alternatively, covariate adjustment can be performed using g‑computation (GC) methods based on an outcome regression (OR) model for the conditional mean of the control outcome given covariate values \citep{z25a,z25b}. There are also doubly robust methods that combine OR and PS models and that retain consistency and asymptotic normality if at least one of the two models is correctly specified \citep{li23,v24,g25,z25a}.

The existing covariate adjustment methods typically assume that control outcomes are exchangeable between internal and external control subjects upon conditioning on a set of baseline covariates that are measured in both the RCT and the external control data. This exchangeability assumption is convenient to use but should not be taken for granted; it can and should be examined by comparing the observed OR patterns in the internal and external control groups \citep{g25,z25b}. To our knowledge, there are mainly two covariate adjustment methods that address possible violations of the exchangeability assumption. One is a selective borrowing method \citep{g25,y25} that allows the exchangeability assumption to be violated by some external control subjects and identifies such violators using the adaptive lasso \citep{g25} or an influence-based procedure \citep{y25}. The other is a GC method \citep{z25b} where possible non-exchangeability is represented as interaction terms in an OR model and the adaptive lasso is used to identify null interaction terms (with zero coefficients). The OR model is assumed to be correctly specified in \citet{z25b}.

In this article, we point out that a particular version of the GC method of \citet{z25b} is protected against misspecification of the OR model. Specifically, if the working OR model is a generalized linear model with a canonical link function and a set of interaction terms involving an external control indicator (to be specified in Section \ref{sec:gc-vs}), the method remains consistent for treatment effect estimation in the RCT population even if the specified OR model is incorrect. This particular method will be referred to as the GC method with variable selection and abbreviated as GC-VS. The GC-VS method inherits an oracle property from the adaptive lasso \citep{z06,lu12} and behaves as if the true set of null interactions were known \textit{a priori}. (Throughout this article, the term \lq\lq interactions" refers specifically to interactions involving the external control indicator in the specified OR model.) If no interactions are null, the GC-VS method is asymptotically equivalent to a standard GC method for covariate adjustment within the RCT \citep{m09,y23,fda23b}. If some interactions are null, the GC-VS method is able to improve efficiency over the GC method based on RCT data alone without introducing an asymptotic bias. If all interactions are null, the GC-VS method is asymptotically equivalent to an existing GC method that incorporates external control data under the assumption of exchangeability \citep{z25a}. These observations hold regardless of the (in)correctness of the working OR model, with the understanding that true parameter values in a misspecified OR model are defined as limits of (unregularized) maximum likelihood estimators.

The rest of the article is organized as follows. In the next section, we set up notations, describe the GC-VS method, present its asymptotic properties, and compare it with other methods. A simulation study is reported in Section \ref{sec:sim}, and an illustrative example given in Section \ref{sec:app}. The article ends with a discussion in Section \ref{sec:disc}.

\section{Methodology}\label{sec:meth}

\subsection{Basic Notations}

For a generic subject in a hybrid control study, let $Z$ be a data source indicator (1 for RCT; 0 for external control), $\bmX$ a vector of baseline covariates, $A$ a treatment indicator (1 for experimental therapy; 0 for control), and $Y$ the clinical outcome of interest. We focus on designs in which only the RCT's control arm is supplemented with external data; that is, $\pr(A=0|Z=0)=1$. The full dataset can be represented as independent observations of $\bmO=(Z,\bmX,A,Y)$, with the $i$-th observation denoted by $\bmO_i=(Z_i,\bmX_i,A_i,Y_i)$, $i=1,\dots,n$.

Our goal is to estimate the effect of the experimental treatment versus control within the RCT population. For each $a\in\{0,1\}$, let $Y(a)$ denote the potential outcome under treatment $a$, and define $\mu_a=\epn\{Y(a)|Z=1\}$ as the mean outcome for treatment $a$ in the RCT population. Common effect measures include the mean difference $\mu_1-\mu_0$, the log mean ratio $\log(\mu_1/\mu_0)$ for outcomes with positive means, and the log odds ratio $\log[\mu_1(1-\mu_0)/\{\mu_0(1-\mu_1)\}]$ for binary outcomes. Each measure can be written as $\delta=g(\mu_1)-g(\mu_0)$, where $g$ is the identity, log, or logit function, respectively. 

\subsection{The GC-VS Method}\label{sec:gc-vs}

In general, GC methods for estimating $(\mu_0,\mu_1,\delta)$ are based on the identity $\mu_a=\epn\{m_a(\bmX)|Z=1\}$, where $m_a(\bmX)=\epn\{Y(a)|Z=1,\bmX\}$, $a=0,1$. Randomization in the RCT implies that $A$ is conditionally independent of $(\bmX,Y(0),Y(1))$ given $Z=1$.
It follows that $m_a(\bmX)=\epn(Y|Z=1,A=a,\bmX)$, $a=0,1$. GC methods take advantage of these relations and estimate each $\mu_a$ as $n_1^{-1}\sum_{i=1}^nZ_i\widehat m_a(\bmX_i)$, where $n_1=\sum_{i=1}^nZ_i$ and $\widehat m_a$ is a generic estimate of $m_a$.

To describe the GC-VS method, we will consider estimating $\mu_0$ first since the external control group provides no new information on $\mu_1$ without making strong assumptions. For estimating $\mu_0$, the GC-VS method aims to borrow information from the external control group in a way that is supported by the data. It starts with a working OR model for the distribution of the control outcome conditional on source and covariates. The model specifies that, given $(A=0,Z,\bmX)$, $Y$ follows a generalized linear model with a canonical link function and with conditional mean
\begin{equation}\label{cont.or.mod}
\epn(Y|A=0,Z,\bmX)=h\left((1,\bmX')\bmbeta+(1-Z)(1,\bmX')\bmgamma\right),
\end{equation}
where $h$ is the inverse link function and $\bmbeta$ and $\bmgamma$ are unknown parameter vectors. The model implies $m_0(\bmX)=h((1,\bmX')\bmbeta)$. The interaction terms, $(1-Z)(1,\bmX')\bmgamma$, play a major role in the GC-VS method and are sometimes referred to as \lq\lq interactions" without qualification. These terms allow the external control group to follow a different OR function than $m_0(\bmX)$. Indeed, equation \eqref{cont.or.mod} can be rewritten as
\begin{equation*}\begin{aligned}
\epn(Y|A=0,Z=1,\bmX)&=h\left((1,\bmX')\bmbeta\right),\\
\epn(Y|A=0,Z=0,\bmX)&=h\left((1,\bmX')\bmbeta_{\text{\sc ec}}\right),
\end{aligned}\end{equation*}
with no assumed relationship between $\bmbeta$ and $\bmbeta_{\text{\sc ec}}=\bmbeta+\bmgamma$, where the subscript EC denotes external control. Without variable selection, model \eqref{cont.or.mod} can be estimated by maximum likelihood. Let $(\widehat\bmbeta^{\text{\sc ml}},\widehat\bmbeta_{\text{\sc ec}}^{\text{\sc ml}})$ be obtained by solving the following likelihood equations:
\begin{equation*}\begin{aligned}
\sum_{i=1}^nZ_i(1-A_i)\left\{Y_i-h\left((1,\bmX_i')\widehat\bmbeta^{\text{\sc ml}}\right)\right\}(1,\bmX_i')'&=\bmzero,\\
\sum_{i=1}^n(1-Z_i)(1-A_i)\left\{Y_i-h\left((1,\bmX_i')\widehat\bmbeta_{\text{\sc ec}}^{\text{\sc ml}}\right)\right\}(1,\bmX_i')'&=\bmzero,
\end{aligned}\end{equation*}
and let $\widehat\bmgamma^{\text{\sc ml}}=\widehat\bmbeta_{\text{\sc ec}}^{\text{\sc ml}}-\widehat\bmbeta^{\text{\sc ml}}$. Note that $\widehat\bmbeta^{\text{\sc ml}}$ is based solely on the RCT data; it does not incorporate any information from the external control data.

A key step in the GC-VS method is to use the adaptive lasso to decide which elements of $\bmgamma$ should be set to 0. Null elements of $\bmgamma$ represent similarities between the internal and external control groups and support information borrowing. Indeed, if some elements of $\bmgamma$ are known to be 0, this knowledge can be used to reduce model \eqref{cont.or.mod} by excluding the null interaction terms, and the reduced model will lead to a restricted maximum likelihood estimator of $\bmbeta$ that is more efficient than $\widehat\bmbeta^{\text{\sc ml}}$, the (unrestricted) maximum likelihood estimator under the \lq\lq full model" \eqref{cont.or.mod}. While knowledge about null elements of $\bmgamma$ is unlikely to be available \textit{a priori}, it can be learned from data using a variable selection method, such as the adaptive lasso \citep{z06}. Write $\bmgamma=(\gamma_1,\dots,\gamma_J)'$ and $\widehat\bmgamma^{\text{\sc ml}}=(\widehat\gamma_1^{\text{\sc ml}},\dots,\widehat\gamma_J^{\text{\sc ml}})'$. The adaptive lasso penalty is given by $\lambda_n\sum_{j=1}^J|\gamma_j/\widehat\gamma_j^{\text{\sc ml}}|$, where $\lambda_n$ is a tuning parameter whose value can be chosen through cross-validation. This penalty term will be subtracted from the log-likelihood for model \eqref{cont.or.mod}, and the penalized log-likelihood will be maximized with respect to $(\bmbeta,\bmgamma)$. Let $(\widehat\bmbeta^{\text{\sc vs}},\widehat\bmgamma^{\text{\sc vs}})$ denote the resulting estimates of $(\bmbeta,\bmgamma)$, which can be found using the R package \texttt{glmnet}. Substituting $\widehat\bmbeta^{\text{\sc vs}}$ into the GC formula leads to
$$
\widehat\mu_0^{\text{\sc gc-vs}}=\frac{1}{n_1}
\sum_{i=1}^nZ_ih\left((1,\bmX_i')\widehat\bmbeta^{\text{\sc vs}}\right).
$$

For estimating $\mu_1$, the GC-VS method makes no attempts to borrow information from the external control data and coincides with a standard GC method based on the RCT data alone. It requires a working OR model for the experimental treatment, separate from the control OR model \eqref{cont.or.mod}. Though unnecessary, it is convenient to specify the experimental OR model as a generalizied linear model similar to \eqref{cont.or.mod}, with the same canonical link function and with conditional mean
\begin{equation}\label{exp.or.mod}
\epn(Y|Z=1,A=1,\bmX)=h\left((1,\bmX')\bmalpha\right),
\end{equation}
where $\bmalpha$ is an unknown parameter vector. Let $\widehat\bmalpha^{\text{\sc ml}}$ denote the maximum likelihood estimate of $\bmalpha$, which solves the equation
$$
\sum_{i=1}^nZ_iA_i\left\{Y_i-h\left((1,\bmX_i')\widehat\bmalpha^{\text{\sc ml}}\right)\right\}(1,\bmX_i')'=\bmzero.
$$
The resulting GC estimator of $\mu_1$ is given by
$$
\widehat\mu_1^{\text{\sc gc-rct}}=\frac{1}{n_1}
\sum_{i=1}^nZ_ih\left((1,\bmX_i')\widehat\bmalpha^{\text{\sc ml}}\right).
$$
Finally, the GC-VS method estimates $\delta=g(\mu_1)-g(\mu_0)$ with
$$
\widehat\delta^{\text{\sc gc-vs}}=g\left(\widehat\mu_1^{\text{\sc gc-rct}}\right)-g\left(\widehat\mu_0^{\text{\sc gc-vs}}\right).
$$

\subsection{Asymptotic Theory}\label{sec:asymp}

It is well known that $\widehat\mu_1^{\text{\sc gc-rct}}$ is consistent for $\mu_1$ and asymptotically normal whether model \eqref{exp.or.mod} is correct or not \citep{m09,z19,z25a}. The asymptotic properties of $\widehat\mu_0^{\text{\sc gc-vs}}$ and $\widehat\delta^{\text{\sc gc-vs}}$ have been studied under the assumption that model \eqref{cont.or.mod} is correctly specified \citep[]{z25b}. Here we provide a more general asymptotic theory for $\widehat\mu_0^{\text{\sc gc-vs}}$ and $\widehat\delta^{\text{\sc gc-vs}}$ that allows model \eqref{cont.or.mod} to be misspecified. This generalization draws upon the theoretical work of \citet{lu12} demonstrating that the adaptive lasso retains its oracle property under certain misspecified models.

Without assuming model \eqref{cont.or.mod} is correct, the \lq\lq true" values of $(\bmbeta,\bmbeta_{\text{\sc ec}},\bmgamma)$ are defined as the limits of $(\widehat\bmbeta^{\text{\sc ml}},\widehat\bmbeta_{\text{\sc ec}}^{\text{\sc ml}},\widehat\bmgamma^{\text{\sc ml}})$ and denoted by $(\bmbeta^*,\bmbeta_{\text{\sc ec}}^*,\bmgamma^*)$. Under mild regularity conditions \citep{v98}, these limits exist and are characterized by the following equations:
\begin{equation*}\begin{aligned}
\epn\left[Z(1-A)\left\{Y-h\left((1,\bmX')\bmbeta^*\right)\right\}(1,\bmX')'\right]&=\bmzero,\\
\epn\left[(1-Z)(1-A)\left\{Y-h\left((1,\bmX')\bmbeta_{\text{\sc ec}}^*\right)\right\}(1,\bmX')'\right]&=\bmzero,\\
\bmgamma^*-\bmbeta_{\text{\sc ec}}^*+\bmbeta^*&=\bmzero.
\end{aligned}\end{equation*}
With $\bmgamma^*=(\gamma_1^*,\dots,\gamma_J^*)'$, define $\mathcal J=\{j:\gamma_j^*\not=0\}$ and $\widehat{\mathcal J}^{\text{\sc vs}}=\{j:\widehat\gamma_j^{\text{\sc vs}}\not=0\}$. Assuming that $\lambda_n\to\infty$ and $n^{-1/2}\lambda_n\to0$ as $n\to\infty$ and that certain regularity conditions hold, Theorem 1 of \citet{lu12} establishes the consistency of variable selection, in the sense that $\pr(\widehat{\mathcal J}^{\text{\sc vs}}=\mathcal J)\to1$, as well as the consistency and asymptotic normality of $\widehat\bmbeta^{\text{\sc vs}}$. Let $\widehat\bmbeta^{\text{oracle}}$ denote the oracle \lq\lq estimator" of $\bmbeta$ obtained from maximum likelihood fitting of the oracle model
$$
\epn(Y|A=0,Z,\bmX)=h\left((1,\bmX')\bmbeta+(1-Z)(1,\bmX')_{\mathcal J}\bmgamma_{\mathcal J}\right),
$$
where the subscript $\mathcal J$ denotes the result of taking a sub-vector with $\mathcal J$ as the index set. Specifically, $\widehat\bmbeta^{\text{oracle}}$ is part of the solution to the oracle likelihood equation
$$
\sum_{i=1}^n(1-A_i)\left\{Y_i-h\left((1,\bmX_i')\widehat\bmbeta^{\text{oracle}}+(1-Z_i)(1,\bmX_i')_{\mathcal J}\widehat\bmgamma_{\mathcal J}^{\text{oracle}}\right)\right\}\left(1,\bmX_i',(1-Z_i)(1,\bmX_i')_{\mathcal J}\right)'=\bmzero.
$$
It follows from the M-estimation theory \citep{v98} that, under standard regularity conditions, $\widehat\bmbeta^{\text{oracle}}$ is consistent for $\bmbeta^*$ and asymptotically linear in the sense that
$$
\sqrt n(\widehat\bmbeta^{\text{oracle}}-\bmbeta^*)=n^{-1/2}\sum_{i=1}^n\bmpsi(\bmO_i)+o_p(1)
$$
for some vector-valued function $\bmpsi$, which is known as the influence function of $\widehat\bmbeta^{\text{oracle}}$. The form of $\bmpsi$ is straightforward to derive but cumbersome to present. According to Theorem 1 of \citet{lu12}, $\widehat\bmbeta^{\text{\sc vs}}$ is also consistent for $\bmbeta^*$, asymptotically linear with the same influence function $\bmpsi$, and therefore asymptotically normal with (scaled) asymptotic variance $\var\{\bmpsi(\bmO)\}$.

The asymptotic properties of $\widehat\mu_0^{\text{\sc gc-vs}}$ and $\widehat\delta^{\text{\sc gc-vs}}$ are provided in the next result, which allows either or both of models \eqref{cont.or.mod} and \eqref{exp.or.mod} to be misspecified. All proofs and regularity conditions are given in Appendix A.

\begin{prop}
Under regularity conditions, we have:
\begin{description}
\item{(a)} $\sqrt n(\widehat\mu_0^{\text{\sc gc-vs}}-\mu_0)$ converges to a normal distribution with mean 0 and variance
$$
\var\left[\frac{Z\{h((1,\bmX')\bmbeta^*)-\mu_0\}}{\tau}+\bmr(\bmbeta^*)'\bmpsi(\bmO)\right],
$$
where $\tau=\pr(Z=1)$, $\bmr(\bmbeta)=\epn\{\dot h((1,\bmX)'\bmbeta)(1,\bmX')'|Z=1\}$, and $\dot h$ is the derivative function of $h$;
\item{(b)} $\sqrt n(\widehat\delta^{\text{\sc gc-vs}}-\delta)$ converges to a normal distribution with mean 0 and variance
\begin{multline*}
\var\Bigg\{\dot g(\mu_1)\left[\frac{Z\{h((1,\bmX')\bmalpha^*)-\mu_1\}}{\tau}+\bmr(\bmalpha^*)'\bmphi(\bmO)\right]\\
-\dot g(\mu_0)\left[\frac{Z\{h((1,\bmX')\bmbeta^*)-\mu_0\}}{\tau}+\bmr(\bmbeta^*)'\bmpsi(\bmO)\right]\Bigg\},
\end{multline*}
where $\dot g$ is the derivative function of $g$, $\bmalpha^*$ is the limit of $\widehat\bmalpha^{\text{\sc ml}}$, and $\bmphi$ is the influence function of $\widehat\bmalpha^{\text{\sc ml}}$.
\end{description}
\end{prop}

For both $\widehat\mu_0^{\text{\sc gc-vs}}$ and $\widehat\delta^{\text{\sc gc-vs}}$, closed-form variance estimates can be obtained by replacing the $\var$ operator with sample variance and unknown quantities with empirical estimates. Alternatively, for ease of implementation, a nonparametric bootstrap procedure can be used to produce variance estimates and confidence intervals.

\subsection{Connections with Other GC Methods}\label{sec:conn}

As a comparator, let us consider a commonly used GC method based on the RCT data alone, which we abbreviate as GC-RCT. The GC-RCT method estimates $\mu_1$ with $\widehat\mu_1^{\text{\sc gc-rct}}$ (as does the GC-VS method), $\mu_0$ with $\widehat\mu_0^{\text{\sc gc-rct}}=n_1^{-1}
\sum_{i=1}^nZ_ih((1,\bmX_i')\widehat\bmbeta^{\text{\sc ml}})$, and $\delta$ with $\widehat\delta^{\text{\sc gc-rct}}=g(\widehat\mu_1^{\text{\sc gc-rct}})-g(\widehat\mu_0^{\text{\sc gc-rct}})$. It differs from the GC-VS method in that $\widehat\mu_0^{\text{\sc gc-rct}}$ is based on $\widehat\bmbeta^{\text{\sc ml}}$ instead of $\widehat\bmbeta^{\text{\sc vs}}$. Defined in Section \ref{sec:gc-vs}, $\widehat\bmbeta^{\text{\sc ml}}$ can be regarded as the maximum likelihood estimate of $\bmbeta$ in the generalized linear model
\begin{equation}\label{cont.or.mod.rct}
\epn(Y|A=0,Z=1,\bmX)=h\left((1,\bmX')\bmbeta\right),
\end{equation}
which results from restricting model \eqref{cont.or.mod} to the internal control group. It is well known that $\widehat\mu_0^{\text{\sc gc-rct}}$ and $\widehat\delta^{\text{\sc gc-rct}}$ remain consistent and asymptotically normal if model \eqref{cont.or.mod.rct} is misspecified \citep{m09,z19,z25a}. Thus, the GC-RCT and GC-VS methods are both model-robust, but they may differ in efficiency, depending on $|\mathcal J|$ (the size of $\mathcal J$) and other factors.

\begin{prop}
Under regularity conditions, we have:
\begin{description}
\item{(a)} If $|\mathcal J|=J$, then $(\widehat\bmbeta^{\text{\sc vs}},\widehat\mu_0^{\text{\sc gc-vs}},\widehat\delta^{\text{\sc gc-vs}})$ are asymptotically equivalent to $(\widehat\bmbeta^{\text{\sc ml}},\widehat\mu_0^{\text{\sc gc-rct}},\widehat\delta^{\text{\sc gc-rct}})$ in the sense of having the same influence functions (and hence the same asymptotic variances);
\item{(b)} If $|\mathcal J|<J$ and model \eqref{cont.or.mod} is correct, then $(\widehat\bmbeta^{\text{\sc vs}},\widehat\mu_0^{\text{\sc gc-vs}})$ are asymptotically more efficient than $(\widehat\bmbeta^{\text{\sc ml}},\widehat\mu_0^{\text{\sc gc-rct}})$;
\item{(c)} If $|\mathcal J|<J$ and models \eqref{cont.or.mod} and \eqref{exp.or.mod} are both correct, then $\widehat\delta^{\text{\sc gc-vs}}$ is asymptotically more efficient than $\widehat\delta^{\text{\sc gc-rct}}$.
\end{description}
\end{prop}

Heuristically, it seems reasonable to expect the GC-VS method, which utilizes the external control data, to be generally more efficient than the GC-RCT method when $|\mathcal J|<J$, even under misspecified models. A definitive theoretical answer to this question is not yet available, but the question will be investigated numerically in a simulation study.

Next, as another comparator, consider a GC method \citep{z25a} based on model \eqref{cont.or.mod} with $\bmgamma$ fixed at $\bmzero$:
\begin{equation}\label{cont.or.mod.ni}
\epn(Y|A=0,Z,\bmX)=h\left((1,\bmX')\bmbeta\right).
\end{equation}
Note that model \eqref{cont.or.mod.ni} applies to both internal and external control subjects and assumes that they satisfy conditional mean exchangeability:
$$
\epn(Y|A=0,Z=0,\bmX)=\epn(Y|A=0,Z=1,\bmX).
$$
Under model \eqref{cont.or.mod.ni}, $\bmbeta$ is estimated by $\widehat\bmbeta_0^{\text{\sc ni}}$, which solves the likelihood equation
$$
\sum_{i=1}^n(1-A_i)\left\{Y_i-h\left((1,\bmX_i')\widehat\bmbeta^{\text{\sc ni}}\right)\right\}(1,\bmX_i')'=\bmzero.
$$
Here, the superscript NI stands for \lq\lq no interactions". The corresponding GC method, abbreviated as GC-NI, estimates $\mu_1$ with $\widehat\mu_1^{\text{\sc gc-rct}}$, $\mu_0$ with $\widehat\mu_0^{\text{\sc gc-ni}}=n_1^{-1}
\sum_{i=1}^nZ_ih((1,\bmX_i')\widehat\bmbeta^{\text{\sc ni}})$, and $\delta$ with $\widehat\delta^{\text{\sc gc-ni}}=g(\widehat\mu_1^{\text{\sc gc-rct}})-g(\widehat\mu_0^{\text{\sc gc-ni}})$.

\begin{prop}
Under regularity conditions, if $|\mathcal J|=0$, then $(\widehat\bmbeta^{\text{\sc vs}},\widehat\mu_0^{\text{\sc gc-vs}},\widehat\delta^{\text{\sc gc-vs}})$ are asymptotically equivalent to $(\widehat\bmbeta^{\text{\sc ni}},\widehat\mu_0^{\text{\sc gc-ni}},\widehat\delta^{\text{\sc gc-ni}})$.
\end{prop}

The GC-NI method was developed and justified under the assumption that model \eqref{cont.or.mod.ni} is correct \citep{z25a}. If model \eqref{cont.or.mod.ni} is correct, then model \eqref{cont.or.mod} is correct with $\bmgamma=\bmgamma^*=\bmzero$. On the other hand, $\bmgamma^*$ can be $\bmzero$ when model \eqref{cont.or.mod.ni} or even model \eqref{cont.or.mod} is misspecified. Thus, Proposition 3 indicates that the GC-NI method is more generally applicable than previously known. However, unlike the GC-VS method, the GC-NI method is not model-robust and may become inconsistent when $|\mathcal J|>0$. To illustrate the last point, Appendix B provides some example scenarios where GC-NI is inconsistent due to an unmeasured prognostic covariate whereas GC-VS remains consistent and potentially more efficient than GC-RCT.

\subsection{Additional Comparative Remarks}

It is of interest to compare the GC-VS method with the selective borrowing method \citep{g25,y25}, the only other method that explicitly addresses non-exchangeability. The two methods target different types of (partial) exchangeability for information borrowing. The GC-VS method exploits null interactions in a specified OR model, while the selective borrowing method operates on a subset of external control subjects (or rather, covariate values) satisfying exchangeability. Example scenarios can be constructed where either or both methods are applicable \citep{z25b}. In terms of modeling assumptions, the selective borrowing method can be implemented parametrically (with parametric OR and PS models) or nonparametrically (using machine learning methods). The parametric version is doubly robust, whereas GC-VS is totally robust, against model misspecification. The nonparametric version of selective borrowing is similar in robustness to GC-VS, although a large sample size may be required for some machine learning methods to perform well.

Some methods incorporate external control data through an established method for covariate adjustment within an RCT. Examples of such methods include the prognostic covariate adjustment (PROCOVA) method \citep{s21}, which estimates a prognostic score using external data and uses the estimated prognostic score as a pre-defined covariate in a covariate-adjusted analysis of the RCT data, and the augmentation method considered by \citet{z25a}, which substitutes an estimate of $m_0(\bmX)$ from external data into an augmentation formula. These methods produce consistent estimators under minimal assumptions but have limited capacity for efficiency improvement. Indeed, while they make use of external control data, their estimation efficiency is subject to the same bound for a covariate-adjusted analysis of the RCT data alone \citep{t08}. This is unsatisfactory because the availability of external control data increases the amount of information and the efficiency bound for treatment effect estimation \citep{li23,v24,g25}. From an efficient estimation point of view, the use of external control data is superficial in the PROCOVA and augmentation methods referenced above.

\section{Simulation}\label{sec:sim}

This section reports a simulation study that evaluates the GC-VS method in comparison with several other methods: the GC-RCT and GC-NI methods described in Section \ref{sec:conn}, two unadjusted (for covariates) methods based on sample averages, and a doubly robust method with selective borrowing (DR-SB) using the adaptive lasso \citep{g25}. One of the unadjusted methods estimates $\mu_a$ with $\{\sum_{i=1}^nZ_iI(A_i=a)\}^{-1}\sum_{i=1}^nZ_iI(A_i=a)Y_i$, where $I(\cdot)$ is the indicator function. This method is based on the RCT data and will be referred to as UA-RCT. The other unadjusted method, abbreviated as UA-pooled, utilizes pooled data and estimates $\mu_a$ with $\{\sum_{i=1}^nI(A_i=a)\}^{-1}\sum_{i=1}^nI(A_i=a)Y_i$. The GC methods are implemented with an identity (for continuous $Y$) or logit (for binary $Y$) link function in models \eqref{cont.or.mod}--\eqref{cont.or.mod.ni}. For all UA and GC methods, analytical standard errors are used to construct confidence intervals. The DR-SB method is implemented using the \texttt{SelectiveIntegrative} package \citep{g25}, with \texttt{method="glm"} and all other options set to default values. Our method comparison will be based on the estimation of $\mu_0$ and $\delta=\mu_1-\mu_0$ as we have not proposed a new estimator of $\mu_1$. For the DR-SB method, the comparison is further limited to the estimation of $\delta$ because the \texttt{SelectiveIntegrative} package does not provide an estimate of $\mu_0$.

We consider two sample size configurations: $n_1=n_0=200$ or 400, where $n_0=n-n_1$ is the size of the external control group. The covariate vector $\bmX=(X_1,X_2,X_3)'$ follows a trivariate normal distribution in each data source. Specifically, given $Z=z\in\{0,1\}$, $\bmX\sim N_3(\bmnu_z,\mbI)$, where $\bmnu_1=\bmzero$, $\bmnu_0=(-0.2, 0.4, 1)'$, and $\mbI$ is the identity matrix. Within the RCT, treatment assignment follows 1:1 randomization (i.e., $\pi=1/2$). For the whole study, the treatment assignment mechanism may be described as $\pr(A=1|Z,\bmX)=\pi Z$. The outcome variable $Y$ may be continuous or binary, and its conditional distribution given $(Z,A,\bmX)$ will be described later in four scenarios. For each sample size configuration and each specified distribution of $(Y|Z,A,\bmX)$, $10^4$ sets of study data are simulated and analyzed using different estimation methods. The only exception here is the DR-SB method, which is computationally demanding and whose evaluation is limited to a random subset of 2000 simulated studies. All other methods are applied to all $10^4$ simulated studies in each case.

In Scenario A, $Y$ is continuous and follows a standard linear regression model:
\begin{equation}\tag{Scenario A}
Y=(1,\bmX')\bmbeta_A+(1-Z)(1,\bmX')\bmgamma_A+\varepsilon,
\end{equation}
where $\bmbeta_A=0.5(1,-1,1,-1)'$ and $\varepsilon\sim N(0,0.2^2)$, independent of $(Z,A,\bmX)$. We consider different choices for $\bmgamma_A$ of the form $(0\bmone_{4-m}',0.75\bmone_m')'$, where $m$ is an integer between 0 and 4 (inclusive) and $\bmone_k$ is a $k$-vector of 1s. This mechanism for generating $Y$ is consistent with models \eqref{cont.or.mod} and \eqref{exp.or.mod} with $h=\text{identity}$, $\bmalpha=\bmbeta=\bmbeta_A$, $\bmgamma=\bmgamma_A$, and $|\mathcal J|=m$. It is also consistent with model \eqref{cont.or.mod.rct} but inconsistent with model \eqref{cont.or.mod.ni} unless $m=0$. Thus, the GC-VS and GC-RCT methods are based on correct working models, while the GC-NI method has a misspecified working model when $m>0$.

Table \ref{sim.rst.A} reports the simulation results in Scenario A in terms of empirical bias, standard deviation, and coverage proportion (at nominal level 95\%). As expected, UA-pooled is severely biased, as is GC-NI when $m>0$, while the other methods exhibit no or negligible bias. Among the (virtually) unbiased methods, GC-RCT is substantially more efficient than UA-RCT, and GC-VS is even more efficient than GC-RCT when $m<4$, whereas DR-SB shows little efficiency improvement over GC-RCT. At $m=0$ (ideal for information borrowing), increasing amounts of efficiency improvement over GC-RCT are observed for DR-SB, GC-VS and GC-NI. For $m\in\{1,2,3\}$, GC-VS attains the highest level of efficiency without introducing bias. For example, comparing GC-VS with GC-RCT for estimating $\delta$ at $n_1=n_0=200$, the reduced standard deviation (0.026 versus 0.029) translates into a 20\% reduction in variance, indicating that the use of GC-VS instead of GC-RCT allows the sample size to be reduced by 20\% while maintaining the same level of estimation precision. When $m=4$, GC-VS and DR-SB show similar efficiency to GC-RCT. Adequate coverage is observed for UA-RCT, GC-RCT and GC-VS at all values of $m$, while the other methods suffer from under-coverage to various degrees. Possible reasons for under-coverage include bias in point estimation for UA-pooled and GC-NI (at $m>0$) and variance under-estimation for DR-SB.

In Scenario B, $Y$ remains continuous but its conditional distribution given $(Z,A,\bmX)$ contains some non-linearity:
\begin{equation}\tag{Scenario B}
Y=(1,\bmX')\bmbeta_B+(1-Z)(1,\bmX')\bmgamma_B+0.5X_1X_2+0.25(X_3^2-1)+\varepsilon,
\end{equation}
where $\bmbeta_B=\bmbeta_A$, $\varepsilon$ is the same as in Scenario A, and $\bmgamma_B$ is chosen such that $\bmgamma^*=\bmgamma_A=(0\bmone_{4-m}',0.75\bmone_m')'$. Specifically, we set
$$
\bmgamma_B=\bmgamma_A-\left[\epn\{(1,\bmX')'(1,\bmX')|Z=0\}\right]^{-1}\epn\left[\{0.5X_1X_2+0.25(X_3^2-1)\}(1,\bmX')'|Z=0\right].
$$
Because of the non-linear terms, this mechanism is clearly inconsistent with models \eqref{cont.or.mod}--\eqref{cont.or.mod.ni}. As a result, all three GC methods are based on incorrect working models. The simulation results in Scenario B, reported in Table \ref{sim.rst.B}, generally follow the same patterns as those in Table \ref{sim.rst.A}, except that the efficiency advantage of GC-RCT over UA-RCT has become smaller. Despite model misspecification, the GC-RCT and GC-VS methods remain unbiased, as does the GC-NI method when $m=0$, as predicted by asymptotic theory. While the asymptotic theory in Section \ref{sec:asymp} does not guarantee an efficiency advantage for GC-VS over GC-RCT with misspecified working models, the efficiency results in Table \ref{sim.rst.B} support the intuition that, by incorporating external control data, GC-VS is likely to improve efficiency over GC-RCT when $m<4$.

In Scenario C, $Y$ is binary and follows a standard logistic regression model:
\begin{equation}\tag{Scenario C}
Y=I\left(U<\expit\big\{(1,\bmX')\bmbeta_C+(1-Z)(1,\bmX')\bmgamma_C\big\}\right),
\end{equation}
where $\expit(u)=1/\{1+\exp(-u)\}$, $\bmbeta_C=\bmbeta_A$, $\bmgamma_C=\bmgamma_A$, and $U$ is uniformly distributed on the unit interval and independent of $(Z,A,\bmX)$. This data generation mechanism is consistent with models \eqref{cont.or.mod}--\eqref{cont.or.mod.rct} with $h=\expit$ but inconsistent with model \eqref{cont.or.mod.ni} unless $m=0$. Thus, as in Scenario A, GC-RCT and GC-VS are based on correct models, whereas GC-NI is based on an incorrect model when $m>0$. The simulation results in Scenario C, shown in Table \ref{sim.rst.C}, are qualitatively similar to the previous results, with a few notable differences. First, at $m=4$, GC-VS exhibits a small bias which diminishes with increasing sample size. Second, while GC-VS is known to be asymptotically equivalent to GC-NI when $m=0$, a large sample size---larger than those considered here---may be required for this asymptotic result to take effect for a binary outcome. Nonetheless, across different values of $m$, GC-VS does maintain a bias advantage over GC-NI and an efficiency advantage over GC-RCT. Third, at $n_1=n_0=200$, GC-VS shows some signs of under-coverage, particularly at $m=4$, but the problem is resolved at $n_1=n_0=400$.

In Scenario D, $Y$ remains binary and is generated as follows:
\begin{equation}\tag{Scenario D}
Y=I\left(U<\expit\big\{(1,\bmX')\bmbeta_D+(1-Z)(1,\bmX')\bmgamma_D+0.5X_1X_2+0.25(X_3^2-1)\big\}\right),
\end{equation}
where $U$ is the same as in Scenario C, $\bmbeta_D=\bmbeta_A$, and $\bmgamma_D$ is chosen such that $\bmgamma^*\approx\bmgamma_A=(0\bmone_{4-m}',0.75\bmone_m')'$. The values of $\bmgamma_D$ are found numerically by analyzing huge sets of simulated data with $n=10^6$. Clearly, this data generation mechanism makes the working models \eqref{cont.or.mod}--\eqref{cont.or.mod.ni} misspecified for all GC methods. Table \ref{sim.rst.D} reports the simulation results in Scenario D, which are quite similar to those in Table \ref{sim.rst.C}.

\section{Application}\label{sec:app}

We now illustrate the methods using a real data example concerning the efficacy of zidovudine (ZDV), an antiretroviral agent that inhibits HIV replication, for treating HIV infection in asymptomatic individuals with hereditary coagulation disorders. This question was examined in an RCT known as ACTG036 \citep{m91}, which enrolled 193 patients and randomized them to ZDV or placebo in a 1:1 ratio. The primary endpoint was the rate of treatment failure, defined as the occurrence of death, acquired immunodeficiency syndrome (AIDS), or advanced AIDS-related complex by 2 years of treatment. Observed failure rates were 4.5\% in the ZDV arm and 7.4\% in the placebo arm, with a difference of $-3.0$\% (95\% CI: $-9.8$\% to 3.9\%). Although the results suggest a potential protective effect of ZDV, the evidence is not definitive.

To bolster the evidence, we incorporate external control data from the placebo arm of ACTG019 \citep{v90}, a randomized trial of ZDV versus placebo for treating HIV infection in asymptomatic patients with CD4 cell count lower than 500/cm$^2$. Patient-level data from both ACTG019 and ACTG036 are publicly available in the R package \texttt{hdbayes}. In ACTG019, the failure rate among the 404 placebo recipients was 8.9\%. Because the two trials enrolled somewhat different patient populations, concerns arise about whether the internal and external control groups are fully comparable. Adjusting for baseline characteristics can help address these differences. The \texttt{hdbayes} version of trial data includes three baseline covariates: age, race (white or non-white), and CD4 cell count.

The data are analyzed using the same six methods compared in Section \ref{sec:sim} with logistic regression models as working models and with the failure rate difference as the effect measure. The covariate vector $\bmX$ consists of age, race, and the square root of CD4 cell count. The same covariate vector is supplied to the DR-SB method. The results of this analysis are reported in Table \ref{hiv.rst.1}, where all three parameters are shown as percentages. To put those results in perspective, we note that UA-pooled and GC-NI are highly susceptible to bias due to their reliance on strong assumptions. Consequently, the results from UA-pooled and GC-NI would not be considered reliable evidence without a suitable justification of the underlying assumptions. Reliable evidence is readily available from UA-RCT and GC-RCT, with GC-RCT being generally more efficient than UA-RCT, but these methods do not incorporate the available external control data. The challenge, then, is how to strengthen the evidence from GC-RCT by incorporating the external control data in a statistically justified manner. This challenge is partially met by DR-SB, which is more robust than GC-NI but still relies on modeling assumptions, and fully met by GC-VS, which is completely model-robust. In this particular example, Table \ref{hiv.rst.1} shows that the GC-VS standard errors for $\delta$ and $\mu_0$ are much smaller than those from GC-RCT and similar to those from GC-NI. This example demonstrates that the GC-VS method can make effective use of external control data without compromising statistical validity.

To illustrate the impact of omitting important covariates, Table \ref{hiv.rst.2} shows the results of a similar analysis with age and race excluded from $\bmX$. The comparison is now limited to GC-RCT, GC-NI and GC-VS because UA-RCT and UA-pooled are indifferent to the choice of $\bmX$ and DR-SB produces no results in this analysis due to a computational issue. The results in Table \ref{hiv.rst.2} are largely similar to those in Table \ref{hiv.rst.1}, with one notable exceptation: the GC-VS standard errors for $\delta$ and $\mu_0$ are now similar to those from GC-RCT, apparently due to the lack of null interactions in model \eqref{cont.or.mod}.

\section{Discussion}\label{sec:disc}

Despite the great potential of the hybrid control design to improve trial efficiency, practitioners are rightfully concerned about its potential to introduce bias into the estimation of treatment effects in the RCT population. Because asymptotically unbiased treatment effect estimators are readily available from the RCT data alone (e.g., GC-RCT), the potential to introduce an asymptotic bias into treatment effect estimation is an undesirable feature for methods that incorporate external control data for improved efficiency. Unlike many of the existing methods for analyzing hybrid control studies, whose consistency relies on strong exchangeability and/or modeling assumptions, the GS-VS method is guaranteed to be consistent under minimal assumptions (i.e., consistency of the adaptive lasso and certain regularity conditions). Simulation results demonstrate that the GC-VS method can effectively improve efficiency over the GC-RCT method without introducing bias at moderate sample sizes. Additionally, the GC-VS method is remarkably simple and easy to implement using standard software (e.g., the R package \texttt{glmnet}). As such, the GC-VS method appears to be a promising approach for analyzing hybrid control studies.

Like all other statistical methods based on asymptotic theory, the GC-VS method faces the practical question of how large the sample size needs to be for the asymptotic theory to take effect. The answer to this question is likely to depend on the investigator's tolerance for bias and under-coverage, the specification and size of the OR model \eqref{cont.or.mod}, the number $|\mathcal J|$ of non-null interactions, and possibly other aspects of the data generation mechanism. In our simulation study, for example, $n_1=n_0=200$ seems adequate in most cases but not when $m=|\mathcal J|=4$ and $Y$ is binary. It seems impractical to identify a minimum sample size that is applicable to all potential applications. A more practical approach is to use simulation experiments to evaluate operating characteristics in the specific context of a given application.

There are some open theoretical questions about the GC-VS method. While the method is known to be asymptotically more efficient than GC-RCT when $|\mathcal J|<J$ and working models are correctly specified, a similar result is not yet available for the more general case where working models may be misspecified. Another pertinent question is how to relax the condition $|\mathcal J|<J$, which requires some components of $\bmgamma^*$ to be exactly 0. In reality, some components of $\bmgamma^*$ may be rather small in absolute value but not exactly 0. To understand the impact of such small values in $\bmgamma^*$, it may be helpful to allow $\bmgamma^*$ to depend on $n$ with some components converging to 0 as $n\to\infty$. Further research on these topics might produce new insights that help to understand or improve the performance of the GC-VS method.







\section*{Appendix A: Proofs}

We assume that models \eqref{cont.or.mod} and \eqref{exp.or.mod} and their likelihood equations satisfy the conditions in Chapter 5 of \citet{v98} that guarantee the existence and uniqueness of $(\bmalpha^*,\bmbeta^*,\bmgamma^*)$ and the consistency and asymptotic linearity of maximum likelihood estimators. We assume that the conditions in Theorem 1 of \citet{lu12} hold for model \eqref{cont.or.mod} so that the adaptive lasso, as applied in Section \ref{sec:gc-vs}, possesses the stated oracle property. We assume that, for some $\epsilon>0$, the classes $\{h((1,\bmX')\bmalpha):\|\bmalpha-\bmalpha^*\|<\epsilon\}$ and $\{h((1,\bmX')\bmbeta):\|\bmbeta-\bmbeta^*\|<\epsilon\}$ are Donsker \citep{vw96} with square-integrable envelopes.  We write $P_0$ for the true distribution of $\bmO$, $P_n$ for the empirical distribution of $\{\bmO_i,i=1,\dots,n\}$, and $Q_n=\sqrt n(P_n-P_0)$ for the empirical process based on the observed data. These will be used as integration operators; for example, we have $\widehat\mu_0^{\text{\sc gc-vs}}=P_n\{Zh((1,\bmX')\widehat\bmbeta^{\text{\sc vs}})\}/P_nZ$.

\subsection*{Proof of Proposition 1}

\subsubsection*{Part (a)}

Clearly, $\widehat\mu_0^{\text{\sc gc-vs}}$ converges in probability to
$$
P_0\{Zh((1,\bmX')\bmbeta^*)\}/P_0Z=\epn\{h((1,\bmX')\bmbeta^*)|Z=1\}.
$$
Although $h((1,\bmX')\bmbeta^*)$ may differ from $m_0(\bmX)$ if model \eqref{cont.or.mod} is misspecified, we will show that $\epn\{h((1,\bmX')\bmbeta^*)|Z=1\}=\mu_0$ without assuming that model \eqref{cont.or.mod} is correctly specified. Recall from Section \ref{sec:asymp} that $\bmbeta^*$ satisfies the equation
$$
\epn\left[Z(1-A)\left\{Y-h\left((1,\bmX')\bmbeta^*\right)\right\}(1,\bmX')'\right]=\bmzero.
$$
The first component of the above equation (corresponding to the \lq\lq intercept") can be re-written as
\begin{equation*}\begin{aligned}
0&=\epn\left[Z(1-A)\left\{Y-h\left((1,\bmX')\bmbeta^*\right)\right\}\right]\\
&=\tau(1-\pi)\epn\left\{Y-h\left((1,\bmX')\bmbeta^*\right)|Z=1,A=0\right\}\\
&=\tau(1-\pi)\left[\epn(Y|Z=1,A=0)-\epn\left\{h\left((1,\bmX')\bmbeta^*\right)|Z=1,A=0\right\}\right]\\
&=\tau(1-\pi)\left[\mu_0-\epn\left\{h((1,\bmX')\bmbeta^*)|Z=1\right\}\right],
\end{aligned}\end{equation*}
where the last step makes use of the conditional independence between $A$ and $\bmX$ given $Z=1$ (due to randomization in the RCT). Because $\tau(1-\pi)\not=0$, we conclude that $\mu_0=\epn\{h((1,\bmX')\bmbeta^*)|Z=1\}$ and that $\widehat\mu_0^{\text{\sc gc-vs}}$ is consistent for $\mu_0$ without assuming model \eqref{cont.or.mod} is correct.

To demonstrate the asymptotic linearity of $\widehat\mu_0^{\text{\sc gc-vs}}$, we write
\begin{equation}\tag{A.1}\begin{aligned}
\sqrt n(\widehat\mu_0^{\text{\sc gc-vs}}-\mu_0)
&=\sqrt n\left[\frac{P_n\{Zh((1,\bmX')\widehat\bmbeta^{\text{\sc vs}})\}}{P_nZ}
-\frac{P_0\{Zh((1,\bmX')\bmbeta^*)\}}{P_0Z}\right]\\
&=\sqrt n\left[\frac{P_n\{Zh((1,\bmX')\widehat\bmbeta^{\text{\sc vs}})\}}{P_nZ}
-\frac{P_n\{Zh((1,\bmX')\widehat\bmbeta^{\text{\sc vs}})\}}{P_0Z}\right]\\
&\quad+\sqrt n\left[\frac{P_n\{Zh((1,\bmX')\widehat\bmbeta^{\text{\sc vs}})\}}{P_0Z}
-\frac{P_0\{Zh((1,\bmX')\widehat\bmbeta^{\text{\sc vs}})\}}{P_0Z}\right]\\
&\quad+\sqrt n\left[\frac{P_0\{Zh((1,\bmX')\widehat\bmbeta^{\text{\sc vs}})\}}{P_0Z}
-\frac{P_0\{Zh((1,\bmX')\bmbeta^*)\}}{P_0Z}\right]\\
&=:D_1+D_2+D_3
\end{aligned}\end{equation}
and analyze the three terms separately. Firstly,
\begin{equation}\tag{A.2}\begin{aligned}
D_1&=\frac{-P_n\{Zh((1,\bmX')\widehat\bmbeta^{\text{\sc vs}})\}Q_nZ}{P_nZP_0Z}
=\frac{-P_0\{Zh((1,\bmX')\bmbeta^*)\}Q_nZ}{P_0ZP_0Z}+o_p(1)\\
&=\frac{-\tau\epn\{h((1,\bmX')\bmbeta^*)|Z=1\}Q_nZ}{\tau^2}+o_p(1)
=\frac{-\mu_0Q_nZ}{\tau}+o_p(1).
\end{aligned}\end{equation}
Secondly, by Lemma 19.24 of \citet{v98},
\begin{equation}\tag{A.3}
D_2=\frac{Q_n\{Zh((1,\bmX')\widehat\bmbeta^{\text{\sc vs}})\}}{P_0Z}
=\frac{Q_n\{Zh((1,\bmX')\bmbeta^*)\}}{\tau}+o_p(1).
\end{equation}
Lastly, by the delta method,
\begin{equation}\tag{A.4}
D_3=\bmr(\bmbeta^*)'\bmpsi(\bmO)+o_p(1),
\end{equation}
where
\begin{equation*}\begin{aligned}
\bmr(\bmbeta)&=\frac{\partial[P_0\{Zh((1,\bmX')\bmbeta)\}/P_0Z]}{\partial\bmbeta}
=\frac{\partial\epn\{h((1,\bmX')\bmbeta)|Z=1\}}{\partial\bmbeta}\\
&=\epn\{\dot h((1,\bmX)'\bmbeta)(1,\bmX')'|Z=1\}.
\end{aligned}\end{equation*}
Substituting (A.2)--(A.4) into (A.1), we obtain
$$
\sqrt n(\widehat\mu_0^{\text{\sc gc-vs}}-\mu_0)
=Q_n\left[\frac{Z\{h((1,\bmX')\bmbeta^*)-\mu_0\}}{\tau}+\bmr(\bmbeta^*)'\bmpsi(\bmO)\right]+o_p(1).
$$

\subsubsection*{Part (b)}

Without assuming that model \eqref{exp.or.mod} is correct, it can be argued as in Part (a) that $\widehat\mu_1^{\text{\sc gc-rct}}$ is consistent for $\mu_1$ and asymptotically linear with
$$
\sqrt n(\widehat\mu_1^{\text{\sc gc-rct}}-\mu_1)=Q_n\left[\frac{Z\{h((1,\bmX')\bmalpha^*)-\mu_1\}}{\tau}+\bmr(\bmalpha^*)'\bmphi(\bmO)\right]+o_p(1).
$$
From this and Part (a), it follows that $\widehat\delta^{\text{\sc gc-vs}}=g(\widehat\mu_1^{\text{\sc gc-rct}})-g(\widehat\mu_0^{\text{\sc gc-vs}})$ is consistent for $\delta$ and asymptotically linear with
\begin{multline*}
\sqrt n(\widehat\delta^{\text{\sc gc-vs}}-\delta)=Q_n\Bigg\{\dot g(\mu_1)\left[\frac{Z\{h((1,\bmX')\bmalpha^*)-\mu_1\}}{\tau}+\bmr(\bmalpha^*)'\bmphi(\bmO)\right]\\
-\dot g(\mu_0)\left[\frac{Z\{h((1,\bmX')\bmbeta^*)-\mu_0\}}{\tau}+\bmr(\bmbeta^*)'\bmpsi(\bmO)\right]\Bigg\}+o_p(1).
\end{multline*}

\subsection*{Proof of Proposition 2}

It is well established that $\widehat\mu_0^{\text{\sc gc-rct}}$ and $\widehat\delta^{\text{\sc gc-rct}}$ are both consistent and asymptotically linear with
\begin{equation*}\begin{aligned}
\sqrt n(\widehat\mu_0^{\text{\sc gc-rct}}-\mu_0)
&=Q_n\left[\frac{Z\{h((1,\bmX')\bmbeta^*)-\mu_0\}}{\tau}+\bmr(\bmbeta^*)'\bmpsi_1(\bmO)\right]+o_p(1),\\
\sqrt n(\widehat\delta^{\text{\sc gc-rct}}-\delta)&=Q_n\Bigg\{\dot g(\mu_1)\left[\frac{Z\{h((1,\bmX')\bmalpha^*)-\mu_1\}}{\tau}+\bmr(\bmalpha^*)'\bmphi(\bmO)\right]\\
&\qquad-\dot g(\mu_0)\left[\frac{Z\{h((1,\bmX')\bmbeta^*)-\mu_0\}}{\tau}+\bmr(\bmbeta^*)'\bmpsi_1(\bmO)\right]\Bigg\}+o_p(1),
\end{aligned}\end{equation*}
where $\bmpsi_1$ is the influence function of $\widehat\bmbeta^{\text{\sc ml}}$.

\subsubsection*{Part (a)}

If $|\mathcal J|=J$, then $\widehat\bmbeta^{\text{\sc ml}}=\widehat\bmbeta^{\text{oracle}}$ and $\bmpsi_1=\bmpsi$, and it follows immediately that $(\widehat\bmbeta^{\text{\sc ml}},\widehat\mu_0^{\text{\sc gc-rct}},\widehat\delta^{\text{\sc gc-rct}})$ have the same influence functions as $(\widehat\bmbeta^{\text{\sc vs}},\widehat\mu_0^{\text{\sc gc-vs}},\widehat\delta^{\text{\sc gc-vs}})$.

\subsubsection*{Part (b)}

Suppose $|\mathcal J|<J$ and model \eqref{cont.or.mod} is correct. In this case, the oracle model
$$
\epn(Y|A=0,Z,\bmX)=h\left((1,\bmX')\bmbeta+(1-Z)(1,\bmX')_{\mathcal J}\bmgamma_{\mathcal J}\right),
$$
is a strict submodel of model \eqref{cont.or.mod}. The oracle estimator $\widehat\bmbeta^{\text{oracle}}$, a maximum likelihood estimator under a correct submodel, is asymptotically more efficient than $\widehat\bmbeta^{\text{\sc ml}}$, the maximum likelihood estimator under the \lq\lq full" model \eqref{cont.or.mod}. Because $\widehat\bmbeta^{\text{\sc vs}}$ is asymptotically equivalent to $\widehat\bmbeta^{\text{oracle}}$, $\widehat\bmbeta^{\text{\sc vs}}$ is also asymptotically more efficient than $\widehat\bmbeta^{\text{\sc ml}}$. Formally, we have $\var\{\bmpsi(\bmO)\}\le\var\{\bmpsi_1(\bmO)\}$ in the sense that $\var\{\bmpsi_1(\bmO)\}-\var\{\bmpsi(\bmO)\}$ is nongenative-definite. The asymptotic variance of $\widehat\mu_0^{\text{\sc gc-vs}}$ is given by
$$
\avar(\widehat\mu_0^{\text{\sc gc-vs}})
=\var\left[\frac{Z\{h((1,\bmX')\bmbeta^*)-\mu_0\}}{\tau}+\bmr(\bmbeta^*)'\bmpsi(\bmO)\right].
$$
It can be shown as in \citet[Appendix A]{z25a} that $\bmpsi(\bmO)$ is uncorrelated with $Z\{\{h((1,\bmX')\bmbeta^*)-\mu_0\}/\tau$ when model \eqref{cont.or.mod} is correct. It follows that
\begin{equation*}\begin{aligned}
\avar(\widehat\mu_0^{\text{\sc gc-vs}})
&=\var\left[\frac{Z\{h((1,\bmX')\bmbeta^*)-\mu_0\}}{\tau}\right]+\var\{\bmr(\bmbeta^*)'\bmpsi(\bmO)\}\\
&=\var\left[\frac{Z\{h((1,\bmX')\bmbeta^*)-\mu_0\}}{\tau}\right]+\bmr(\bmbeta^*)'\var\{\bmpsi(\bmO)\}\bmr(\bmbeta^*).
\end{aligned}\end{equation*}
Similarly, the asymptotic variance of $\widehat\mu_0^{\text{\sc gc-rct}}$ is found to be
$$
\avar(\widehat\mu_0^{\text{\sc gc-rct}})
=\var\left[\frac{Z\{h((1,\bmX')\bmbeta^*)-\mu_0\}}{\tau}\right]+\bmr(\bmbeta^*)'\var\{\bmpsi_1(\bmO)\}\bmr(\bmbeta^*).
$$
Because $\var\{\bmpsi(\bmO)\}\le\var\{\bmpsi_1(\bmO)\}$, we conclude that $\avar(\widehat\mu_0^{\text{\sc gc-vs}})\le\avar(\widehat\mu_0^{\text{\sc gc-rct}})$.

\subsubsection*{Part (c)}

Suppose $|\mathcal J|<J$ and models \eqref{cont.or.mod} and \eqref{exp.or.mod} are both correct. The asymptotic variance of $\widehat\delta^{\text{\sc gc-vs}}$ is given by
$$
\avar(\widehat\delta^{\text{\sc gc-vs}})=\var\left\{B+\dot g(\mu_1)\bmr(\bmalpha^*)'\bmphi(\bmO)
-\dot g(\mu_0)\bmr(\bmbeta^*)'\bmpsi(\bmO)\right\},
$$
where
$$
B=\tau^{-1}Z\left[\dot g(\mu_1)\{h((1,\bmX')\bmalpha^*)-\mu_1\}-\dot g(\mu_0)\{h((1,\bmX')\bmbeta^*)-\mu_0\}\right].
$$
It can be argued as in \citet[Appendix A]{z25a} that $B$ is uncorrelated with both $\bmphi(\bmO)$ and $\bmpsi(\bmO)$ when models \eqref{cont.or.mod} and \eqref{exp.or.mod} are both correct. Furthermore, because $\bmphi(\bmO)$ is a multiple of $A$ and $\bmpsi(\bmO)$ is a multiple of $(1-A)$, they are uncorrelated with each other. It follows that
$$
\avar(\widehat\delta^{\text{\sc gc-vs}})=\var(B)+\dot g(\mu_1)^2\bmr(\bmalpha^*)'\var\{\bmphi(\bmO)\}\bmr(\bmalpha^*)+\dot g(\mu_0)^2\bmr(\bmbeta^*)'\var\{\bmpsi(\bmO)\}\bmr(\bmbeta^*).
$$
Similarly, the asymptotic variance of $\widehat\delta^{\text{\sc gc-rct}}$ is found to be
$$
\avar(\widehat\delta^{\text{\sc gc-rct}})=\var(B)+\dot g(\mu_1)^2\bmr(\bmalpha^*)'\var\{\bmphi(\bmO)\}\bmr(\bmalpha^*)+\dot g(\mu_0)^2\bmr(\bmbeta^*)'\var\{\bmpsi_1(\bmO)\}\bmr(\bmbeta^*).
$$
Because $\var\{\bmpsi(\bmO)\}\le\var\{\bmpsi_1(\bmO)\}$, we conclude that $\avar(\widehat\delta^{\text{\sc gc-vs}})\le\avar(\widehat\delta^{\text{\sc gc-rct}})$.

\subsection*{Proof of Proposition 3}

Suppose $|\mathcal J|=0$. In this case, $\widehat\bmbeta^{\text{\sc ni}}$ is identical to $\widehat\bmbeta^{\text{oracle}}$, which is asymptotically equivalent to $\widehat\bmbeta^{\text{\sc vs}}$. In particular, $\widehat\bmbeta^{\text{\sc ni}}$ is consistent for $\bmbeta^*$ and asymptotically linear with influence function $\bmpsi$. Based on this fact, it can be shown as in the proof of Proposition 1 that $(\widehat\mu_0^{\text{\sc gc-ni}},\widehat\delta^{\text{\sc gc-ni}})$ are consistent for $(\mu_0,\delta)$ and asymptotically linear with the same influence functions as $(\widehat\mu_0^{\text{\sc gc-vs}},\widehat\delta^{\text{\sc gc-vs}})$.

\section*{Appendix B: Impact of Unmeasured Covariates}

Suppose a continuous outcome is to be analyzed using linear regression models, i.e., models \eqref{cont.or.mod}--\eqref{cont.or.mod.ni} with $h=\text{identity}$. Let us imagine that there is an unmeasured baseline variable $V$ which, together with $\bmX$, makes the control outcome mean-exchangeable between the internal and external control groups:
$$
\epn(Y|A=0,Z=0,\bmX,V)=\epn(Y|A=0,Z=1,\bmX,V).
$$
For concreteness and simplicity, suppose the above conditional expectations are linear in $\bmX$ and $V$:
\begin{equation}\tag{B.1}
\epn(Y|A=0,Z,\bmX,V)=(1,\bmX')\bmb+V
\end{equation}
for some vector $\bmb$.

To determine the (in)consistency of the GC-NI method (with $h=\text{identity}$), it is necessary to consider whether $\epn(Y|A=0,Z,\bmX)$ (without conditioning on $V$) is correctly described by model \eqref{cont.or.mod.ni}. To this end, we write
\begin{equation*}\begin{aligned}
\epn(Y|A=0,Z,\bmX)&=\epn\big\{\epn(Y|A=0,Z,\bmX,V)\big|A=0,Z,\bmX\big\}\\
&=\epn\big\{(1,\bmX')\bmb+V\big|A=0,Z,\bmX\big\}\\
&=(1,\bmX')\bmb+\epn(V|A=0,Z,\bmX)\\
&=(1,\bmX')\bmb+\epn(V|Z,\bmX),
\end{aligned}\end{equation*}
where the first step follows from the law of iterated expectations, the second one from (B.1), and the last one from the hybrid control design, which implies that $A$ is conditionally independent of $(\bmX,V)$ given $Z$. Comparing the preceding display with model \eqref{cont.or.mod.ni}, it is easy to see that, for the GC-NI method to be consistent, $\epn(V|Z,\bmX)$ must not depend on $Z$ and must be linear in $\bmX$, that is,
\begin{equation}\tag{B.2}
\epn(V|Z,\bmX)=(1,\bmX')\bmc
\end{equation}
for some vector $\bmc$. If this is not the case, model \eqref{cont.or.mod.ni} will be misspecified and the GC-NI method inconsistent.

In contrast, by Proposition 1, the GC-VS method remains consistent regardless of the (in)correctness of model \eqref{cont.or.mod}. In that regard, the GC-VS method is indifferent to the existence of $V$ as well as the specific forms of $\epn(Y|A=0,Z,\bmX,V)$ and $\epn(V|Z,\bmX)$. By proposition 2, the GC-VS method will be more efficient than the GC-RCT method, which is also model-robust, if model \eqref{cont.or.mod} is correct and $\bmgamma^*$ has one or more null elements. This can happen in the presence of $V$, under weaker conditions than those required for GC-NI to be consistent. For example, under (B.1) and (B.2), $\bmgamma^*$ is entirely null, and it follows from Proposition 3 that GC-VS is asymptotically equivalent to GC-NI, both being more efficient than GC-RCT. In fact, for GC-VS to maintain an efficiency advantage over GC-RCT, (B.2) can be relaxed as follows:
\begin{equation}\tag{B.3}
\epn(V|Z,\bmX)=(1,\bmX')\bmc+(1-Z)(1,\bmX')\bmd,
\end{equation}
where $\bmd$ is another vector of the same dimension as $\bmb$ and $\bmc$. The additional terms in (B.3) relative to (B.2) allows $V$ to be differentially distributed between the trial population ($Z=1$) and the external control population ($Z=0$) after adjusting for $\bmX$. (B.1) and (B.3) together imply that
$$
\epn(Y|A=0,Z,\bmX)=(1,\bmX')(\bmb+\bmc)+(1-Z)(1,\bmX')\bmd,
$$
which further implies that model \eqref{cont.or.mod} is correct with $\bmbeta^*=\bmb+\bmc$ and $\bmgamma^*=\bmd$. Thus, by Proposition 2, GC-VS will be more efficient than GC-RCT if $\bmd$ has at least one null element.

\pagebreak
\begin{landscape}
\renewcommand{\baselinestretch}{1.0}
\begin{table}[htbp]
{\footnotesize
\caption{Simulation results in Scenario A (continuous outcome, correct working models): empirical bias, standard deviation (SD), and coverage proportion (CP) for estimating $(\mu_0,\delta)$ using six different estimation methods (see Section \ref{sec:sim} for details).}\label{sim.rst.A}
\newcolumntype{d}{D{.}{.}{3}}
\begin{center}
\begin{tabular}{clddcddcddcddcddcddc}
\hline
\hline
&&\multicolumn{8}{c}{$n_1=n_0=200$}&&\multicolumn{8}{c}{$n_1=n_0=400$}\\
\cline{3-10}\cline{12-19}
$m=|\mathcal J|$&\multicolumn{1}{c}{Method}&\multicolumn{2}{c}{Bias}&&\multicolumn{2}{c}{SD}&&\multicolumn{2}{c}{CP}&&\multicolumn{2}{c}{Bias}&&\multicolumn{2}{c}{SD}&&\multicolumn{2}{c}{CP}\\
\cline{3-4}\cline{6-7}\cline{9-10}\cline{12-13}\cline{15-16}\cline{18-19}
&&\multicolumn{1}{c}{$\mu_0$}&\multicolumn{1}{c}{$\delta$}&&\multicolumn{1}{c}{$\mu_0$}&\multicolumn{1}{c}{$\delta$}&&\multicolumn{1}{c}{$\mu_0$}&\multicolumn{1}{c}{$\delta$}&&\multicolumn{1}{c}{$\mu_0$}&\multicolumn{1}{c}{$\delta$}&&\multicolumn{1}{c}{$\mu_0$}&\multicolumn{1}{c}{$\delta$}&&\multicolumn{1}{c}{$\mu_0$}&\multicolumn{1}{c}{$\delta$}\\
\hline
0&UA-RCT&0.005&-0.012&&0.094&0.126&&0.929&0.947&&0.002&-0.004&&0.063&0.088&&0.955&0.949\\
&UA-pooled&-0.134&0.127&&0.054&0.103&&0.253&0.762&&-0.131&0.129&&0.037&0.071&&0.052&0.561\\
&GC-RCT&0.000&-0.002&&0.068&0.029&&0.947&0.949&&0.000&0.000&&0.044&0.020&&0.953&0.951\\
&GC-NI&-0.001&-0.001&&0.066&0.026&&0.943&0.933&&0.000&0.000&&0.043&0.017&&0.948&0.948\\
&GC-VS&-0.001&-0.002&&0.067&0.027&&0.941&0.939&&0.000&0.000&&0.044&0.018&&0.949&0.953\\
&DR-SB&&0.002&&&0.028&&&0.877&&&-0.001&&&0.019&&&0.871\\
\hline
1&UA-RCT&0.004&-0.012&&0.094&0.126&&0.930&0.947&&0.003&-0.006&&0.062&0.087&&0.958&0.949\\
&UA-pooled&0.368&-0.376&&0.050&0.101&&0.000&0.036&&0.369&-0.372&&0.034&0.070&&0.000&0.000\\
&GC-RCT&0.000&-0.002&&0.067&0.029&&0.947&0.949&&0.000&0.000&&0.044&0.019&&0.954&0.952\\
&GC-NI&0.132&-0.134&&0.063&0.043&&0.408&0.107&&0.132&-0.132&&0.041&0.028&&0.107&0.006\\
&GC-VS&0.001&-0.003&&0.065&0.026&&0.941&0.935&&0.002&-0.002&&0.043&0.017&&0.949&0.949\\
&DR-SB&&0.002&&&0.029&&&0.876&&&-0.002&&&0.019&&&0.915\\
\hline
2&UA-RCT&0.004&-0.012&&0.094&0.127&&0.929&0.946&&0.003&-0.006&&0.062&0.087&&0.958&0.949\\
&UA-pooled&0.568&-0.576&&0.076&0.115&&0.000&0.000&&0.571&-0.574&&0.051&0.080&&0.000&0.000\\
&GC-RCT&0.000&-0.002&&0.068&0.029&&0.947&0.949&&0.000&0.000&&0.044&0.019&&0.954&0.953\\
&GC-NI&0.184&-0.187&&0.082&0.054&&0.328&0.067&&0.185&-0.185&&0.052&0.035&&0.063&0.000\\
&GC-VS&0.002&-0.004&&0.066&0.026&&0.943&0.935&&0.003&-0.003&&0.043&0.017&&0.952&0.949\\
&DR-SB&&0.001&&&0.028&&&0.905&&&-0.002&&&0.019&&&0.896\\
\hline
3&UA-RCT&0.004&-0.012&&0.094&0.126&&0.930&0.947&&0.003&-0.006&&0.062&0.087&&0.958&0.949\\
&UA-pooled&0.469&-0.477&&0.071&0.111&&0.000&0.008&&0.471&-0.474&&0.048&0.077&&0.000&0.000\\
&GC-RCT&0.000&-0.002&&0.067&0.029&&0.947&0.949&&0.000&0.000&&0.044&0.019&&0.954&0.952\\
&GC-NI&0.160&-0.162&&0.077&0.060&&0.410&0.238&&0.159&-0.159&&0.050&0.040&&0.120&0.027\\
&GC-VS&0.001&-0.003&&0.066&0.026&&0.941&0.937&&0.002&-0.002&&0.043&0.017&&0.951&0.953\\
&DR-SB&&0.001&&&0.028&&&0.886&&&-0.002&&&0.019&&&0.905\\
\hline
4&UA-RCT&0.004&-0.012&&0.094&0.126&&0.930&0.947&&0.003&-0.006&&0.062&0.088&&0.958&0.949\\
&UA-pooled&0.969&-0.977&&0.074&0.113&&0.000&0.000&&0.971&-0.974&&0.050&0.078&&0.000&0.000\\
&GC-RCT&0.000&-0.002&&0.067&0.029&&0.947&0.949&&0.000&0.000&&0.044&0.019&&0.954&0.953\\
&GC-NI&0.552&-0.554&&0.080&0.066&&0.000&0.000&&0.553&-0.553&&0.052&0.044&&0.000&0.000\\
&GC-VS&0.004&-0.006&&0.067&0.029&&0.943&0.937&&0.004&-0.004&&0.044&0.019&&0.951&0.945\\
&DR-SB&&0.001&&&0.029&&&0.867&&&-0.002&&&0.019&&&0.914\\
\hline
\end{tabular}
\end{center}
}
\end{table}
\end{landscape}

\pagebreak
\begin{landscape}
\renewcommand{\baselinestretch}{1.0}
\begin{table}[htbp]
{\footnotesize
\caption{Simulation results in Scenario B (continuous outcome, incorrect working models): empirical bias, standard deviation (SD), and coverage proportion (CP) for estimating $(\mu_0,\delta)$ using six different estimation methods (see Section \ref{sec:sim} for details).}\label{sim.rst.B}
\newcolumntype{d}{D{.}{.}{3}}
\begin{center}
\begin{tabular}{clddcddcddcddcddcddc}
\hline
\hline
&&\multicolumn{8}{c}{$n_1=n_0=200$}&&\multicolumn{8}{c}{$n_1=n_0=400$}\\
\cline{3-10}\cline{12-19}
$m=|\mathcal J|$&\multicolumn{1}{c}{Method}&\multicolumn{2}{c}{Bias}&&\multicolumn{2}{c}{SD}&&\multicolumn{2}{c}{CP}&&\multicolumn{2}{c}{Bias}&&\multicolumn{2}{c}{SD}&&\multicolumn{2}{c}{CP}\\
\cline{3-4}\cline{6-7}\cline{9-10}\cline{12-13}\cline{15-16}\cline{18-19}
&&\multicolumn{1}{c}{$\mu_0$}&\multicolumn{1}{c}{$\delta$}&&\multicolumn{1}{c}{$\mu_0$}&\multicolumn{1}{c}{$\delta$}&&\multicolumn{1}{c}{$\mu_0$}&\multicolumn{1}{c}{$\delta$}&&\multicolumn{1}{c}{$\mu_0$}&\multicolumn{1}{c}{$\delta$}&&\multicolumn{1}{c}{$\mu_0$}&\multicolumn{1}{c}{$\delta$}&&\multicolumn{1}{c}{$\mu_0$}&\multicolumn{1}{c}{$\delta$}\\
\hline
0&UA-RCT&0.003&-0.007&&0.110&0.155&&0.941&0.945&&0.002&-0.007&&0.076&0.106&&0.955&0.955\\
&UA-pooled&-0.135&0.130&&0.063&0.123&&0.411&0.808&&-0.131&0.127&&0.045&0.086&&0.156&0.698\\
&GC-RCT&-0.002&-0.002&&0.090&0.092&&0.945&0.941&&-0.002&-0.002&&0.061&0.064&&0.954&0.945\\
&GC-NI&-0.002&-0.002&&0.085&0.082&&0.942&0.941&&0.000&-0.003&&0.057&0.057&&0.953&0.949\\
&GC-VS&-0.003&-0.002&&0.085&0.085&&0.949&0.942&&-0.001&-0.002&&0.058&0.059&&0.954&0.946\\
&DR-SB&&0.001&&&0.093&&&0.856&&&-0.011&&&0.069&&&0.836\\
\hline
1&UA-RCT&0.006&-0.013&&0.111&0.156&&0.939&0.945&&0.002&-0.007&&0.076&0.106&&0.954&0.956\\
&UA-pooled&0.367&-0.374&&0.061&0.121&&0.000&0.124&&0.368&-0.373&&0.042&0.084&&0.000&0.010\\
&GC-RCT&-0.001&-0.003&&0.090&0.092&&0.945&0.937&&-0.002&-0.002&&0.062&0.064&&0.954&0.945\\
&GC-NI&0.131&-0.135&&0.083&0.089&&0.617&0.657&&0.132&-0.135&&0.056&0.060&&0.345&0.398\\
&GC-VS&0.003&-0.007&&0.085&0.087&&0.949&0.935&&0.000&-0.004&&0.059&0.060&&0.945&0.948\\
&DR-SB&&0.003&&&0.093&&&0.867&&&-0.013&&&0.070&&&0.827\\
\hline
2&UA-RCT&0.006&-0.013&&0.111&0.156&&0.939&0.945&&0.002&-0.007&&0.076&0.106&&0.954&0.956\\
&UA-pooled&0.567&-0.574&&0.084&0.133&&0.000&0.015&&0.570&-0.575&&0.056&0.091&&0.000&0.000\\
&GC-RCT&-0.001&-0.004&&0.090&0.092&&0.945&0.937&&-0.002&-0.002&&0.062&0.064&&0.954&0.945\\
&GC-NI&0.183&-0.187&&0.097&0.094&&0.512&0.476&&0.185&-0.188&&0.065&0.065&&0.194&0.179\\
&GC-VS&0.002&-0.007&&0.087&0.087&&0.945&0.939&&0.001&-0.004&&0.059&0.060&&0.953&0.945\\
&DR-SB&&0.005&&&0.092&&&0.866&&&-0.012&&&0.069&&&0.848\\
\hline
3&UA-RCT&0.006&-0.013&&0.111&0.156&&0.939&0.945&&0.002&-0.007&&0.076&0.106&&0.954&0.956\\
&UA-pooled&0.468&-0.475&&0.079&0.131&&0.000&0.051&&0.470&-0.475&&0.053&0.088&&0.000&0.000\\
&GC-RCT&-0.001&-0.004&&0.090&0.092&&0.945&0.937&&-0.002&-0.002&&0.062&0.064&&0.954&0.945\\
&GC-NI&0.158&-0.163&&0.093&0.098&&0.577&0.602&&0.158&-0.162&&0.063&0.067&&0.295&0.349\\
&GC-VS&0.000&-0.004&&0.086&0.088&&0.945&0.931&&0.000&-0.004&&0.059&0.061&&0.949&0.945\\
&DR-SB&&0.007&&&0.092&&&0.867&&&-0.010&&&0.069&&&0.858\\
\hline
4&UA-RCT&0.006&-0.013&&0.111&0.156&&0.939&0.945&&0.002&-0.007&&0.076&0.106&&0.954&0.956\\
&UA-pooled&0.968&-0.975&&0.082&0.132&&0.000&0.000&&0.970&-0.975&&0.054&0.089&&0.000&0.000\\
&GC-RCT&-0.001&-0.003&&0.090&0.092&&0.945&0.937&&-0.002&-0.002&&0.062&0.064&&0.954&0.945\\
&GC-NI&0.551&-0.555&&0.096&0.102&&0.000&0.000&&0.552&-0.556&&0.066&0.070&&0.000&0.000\\
&GC-VS&0.005&-0.009&&0.091&0.093&&0.941&0.939&&0.002&-0.006&&0.062&0.064&&0.947&0.943\\
&DR-SB&&0.004&&&0.093&&&0.876&&&-0.011&&&0.069&&&0.858\\
\hline
\end{tabular}
\end{center}
}
\end{table}
\end{landscape}

\pagebreak
\begin{landscape}
\renewcommand{\baselinestretch}{1.0}
\begin{table}[htbp]
{\footnotesize
\caption{Simulation results in Scenario C (binary outcome, correct working models): empirical bias, standard deviation (SD), and coverage proportion (CP) for estimating $(\mu_0,\delta)$ using six different estimation methods (see Section \ref{sec:sim} for details).}\label{sim.rst.C}
\newcolumntype{d}{D{.}{.}{3}}
\begin{center}
\begin{tabular}{clddcddcddcddcddcddc}
\hline
\hline
&&\multicolumn{8}{c}{$n_1=n_0=200$}&&\multicolumn{8}{c}{$n_1=n_0=400$}\\
\cline{3-10}\cline{12-19}
$m=|\mathcal J|$&\multicolumn{1}{c}{Method}&\multicolumn{2}{c}{Bias}&&\multicolumn{2}{c}{SD}&&\multicolumn{2}{c}{CP}&&\multicolumn{2}{c}{Bias}&&\multicolumn{2}{c}{SD}&&\multicolumn{2}{c}{CP}\\
\cline{3-4}\cline{6-7}\cline{9-10}\cline{12-13}\cline{15-16}\cline{18-19}
&&\multicolumn{1}{c}{$\mu_0$}&\multicolumn{1}{c}{$\delta$}&&\multicolumn{1}{c}{$\mu_0$}&\multicolumn{1}{c}{$\delta$}&&\multicolumn{1}{c}{$\mu_0$}&\multicolumn{1}{c}{$\delta$}&&\multicolumn{1}{c}{$\mu_0$}&\multicolumn{1}{c}{$\delta$}&&\multicolumn{1}{c}{$\mu_0$}&\multicolumn{1}{c}{$\delta$}&&\multicolumn{1}{c}{$\mu_0$}&\multicolumn{1}{c}{$\delta$}\\
\hline
0&UA-RCT&0.000&0.000&&0.049&0.070&&0.945&0.948&&0.000&0.000&&0.035&0.049&&0.949&0.950\\
&UA-pooled&-0.028&0.028&&0.028&0.057&&0.839&0.914&&-0.028&0.028&&0.020&0.040&&0.727&0.893\\
&GC-RCT&0.000&-0.001&&0.048&0.065&&0.935&0.946&&0.000&0.000&&0.034&0.046&&0.947&0.948\\
&GC-NI&0.000&0.000&&0.034&0.056&&0.949&0.946&&0.000&0.000&&0.023&0.039&&0.951&0.951\\
&GC-VS&0.000&0.000&&0.040&0.059&&0.934&0.945&&0.000&0.000&&0.028&0.042&&0.945&0.946\\
&DR-SB&&-0.005&&&0.064&&&0.886&&&-0.001&&&0.046&&&0.889\\
\hline
1&UA-RCT&0.000&0.000&&0.049&0.070&&0.945&0.948&&0.000&0.000&&0.035&0.049&&0.949&0.949\\
&UA-pooled&0.076&-0.076&&0.027&0.056&&0.207&0.734&&0.075&-0.075&&0.019&0.039&&0.027&0.525\\
&GC-RCT&0.000&0.000&&0.048&0.065&&0.935&0.946&&0.000&0.000&&0.034&0.046&&0.946&0.947\\
&GC-NI&0.030&-0.031&&0.034&0.056&&0.832&0.905&&0.031&-0.031&&0.024&0.039&&0.751&0.879\\
&GC-VS&0.005&-0.005&&0.043&0.061&&0.923&0.940&&0.002&-0.003&&0.030&0.043&&0.937&0.944\\
&DR-SB&&-0.006&&&0.064&&&0.883&&&0.000&&&0.047&&&0.870\\
\hline
2&UA-RCT&0.000&0.000&&0.049&0.070&&0.945&0.948&&0.000&0.000&&0.035&0.049&&0.948&0.949\\
&UA-pooled&0.085&-0.085&&0.026&0.056&&0.121&0.676&&0.085&-0.085&&0.019&0.039&&0.009&0.425\\
&GC-RCT&0.000&-0.001&&0.048&0.065&&0.935&0.946&&0.000&0.000&&0.034&0.046&&0.946&0.948\\
&GC-NI&0.028&-0.029&&0.034&0.056&&0.855&0.918&&0.028&-0.029&&0.024&0.038&&0.773&0.895\\
&GC-VS&0.003&-0.004&&0.044&0.062&&0.926&0.942&&0.002&-0.002&&0.031&0.044&&0.936&0.945\\
&DR-SB&&-0.006&&&0.064&&&0.893&&&0.000&&&0.046&&&0.881\\
\hline
3&UA-RCT&0.000&0.000&&0.049&0.070&&0.945&0.948&&0.000&0.000&&0.035&0.049&&0.949&0.949\\
&UA-pooled&0.072&-0.072&&0.027&0.057&&0.239&0.747&&0.072&-0.072&&0.019&0.039&&0.039&0.554\\
&GC-RCT&0.000&0.000&&0.048&0.065&&0.935&0.945&&0.000&0.000&&0.034&0.046&&0.947&0.948\\
&GC-NI&0.024&-0.025&&0.034&0.056&&0.874&0.920&&0.024&-0.025&&0.024&0.039&&0.812&0.907\\
&GC-VS&0.002&-0.003&&0.044&0.062&&0.933&0.942&&0.000&-0.001&&0.031&0.044&&0.937&0.939\\
&DR-SB&&-0.006&&&0.065&&&0.898&&&0.000&&&0.047&&&0.876\\
\hline
4&UA-RCT&0.000&0.000&&0.049&0.070&&0.945&0.948&&0.000&0.000&&0.035&0.049&&0.949&0.949\\
&UA-pooled&0.142&-0.142&&0.025&0.055&&0.001&0.265&&0.142&-0.142&&0.017&0.038&&0.000&0.035\\
&GC-RCT&0.000&-0.001&&0.048&0.065&&0.935&0.946&&0.000&0.000&&0.034&0.046&&0.946&0.948\\
&GC-NI&0.088&-0.088&&0.033&0.056&&0.248&0.639&&0.088&-0.088&&0.023&0.038&&0.033&0.384\\
&GC-VS&0.009&-0.009&&0.051&0.067&&0.890&0.925&&0.003&-0.003&&0.035&0.046&&0.934&0.941\\
&DR-SB&&-0.006&&&0.064&&&0.892&&&-0.001&&&0.047&&&0.866\\
\hline
\end{tabular}
\end{center}
}
\end{table}
\end{landscape}

\pagebreak
\begin{landscape}
\renewcommand{\baselinestretch}{1.0}
\begin{table}[htbp]
{\footnotesize
\caption{Simulation results in Scenario D (binary outcome, incorrect working models): empirical bias, standard deviation (SD), and coverage proportion (CP) for estimating $(\mu_0,\delta)$ using six different estimation methods (see Section \ref{sec:sim} for details).}\label{sim.rst.D}
\newcolumntype{d}{D{.}{.}{3}}
\begin{center}
\begin{tabular}{clddcddcddcddcddcddc}
\hline
\hline
&&\multicolumn{8}{c}{$n_1=n_0=200$}&&\multicolumn{8}{c}{$n_1=n_0=400$}\\
\cline{3-10}\cline{12-19}
$m=|\mathcal J|$&\multicolumn{1}{c}{Method}&\multicolumn{2}{c}{Bias}&&\multicolumn{2}{c}{SD}&&\multicolumn{2}{c}{CP}&&\multicolumn{2}{c}{Bias}&&\multicolumn{2}{c}{SD}&&\multicolumn{2}{c}{CP}\\
\cline{3-4}\cline{6-7}\cline{9-10}\cline{12-13}\cline{15-16}\cline{18-19}
&&\multicolumn{1}{c}{$\mu_0$}&\multicolumn{1}{c}{$\delta$}&&\multicolumn{1}{c}{$\mu_0$}&\multicolumn{1}{c}{$\delta$}&&\multicolumn{1}{c}{$\mu_0$}&\multicolumn{1}{c}{$\delta$}&&\multicolumn{1}{c}{$\mu_0$}&\multicolumn{1}{c}{$\delta$}&&\multicolumn{1}{c}{$\mu_0$}&\multicolumn{1}{c}{$\delta$}&&\multicolumn{1}{c}{$\mu_0$}&\multicolumn{1}{c}{$\delta$}\\
\hline
0&UA-RCT&-0.001&0.002&&0.049&0.070&&0.947&0.945&&0.000&0.000&&0.034&0.048&&0.952&0.955\\
&UA-pooled&-0.018&0.019&&0.029&0.057&&0.899&0.928&&-0.018&0.018&&0.020&0.039&&0.850&0.926\\
&GC-RCT&-0.001&0.001&&0.048&0.066&&0.943&0.941&&0.000&0.000&&0.033&0.045&&0.949&0.956\\
&GC-NI&-0.001&0.001&&0.033&0.056&&0.945&0.944&&0.000&0.000&&0.023&0.038&&0.945&0.959\\
&GC-VS&-0.001&0.001&&0.040&0.060&&0.935&0.935&&0.000&0.000&&0.028&0.041&&0.937&0.951\\
&DR-SB&&-0.001&&&0.065&&&0.881&&&0.001&&&0.046&&&0.888\\
\hline
1&UA-RCT&-0.001&0.002&&0.049&0.070&&0.948&0.945&&0.000&0.000&&0.034&0.047&&0.953&0.956\\
&UA-pooled&0.083&-0.082&&0.027&0.057&&0.141&0.695&&0.083&-0.083&&0.019&0.038&&0.010&0.440\\
&GC-RCT&-0.002&0.002&&0.048&0.066&&0.944&0.941&&0.000&0.000&&0.033&0.045&&0.951&0.956\\
&GC-NI&0.029&-0.029&&0.034&0.057&&0.842&0.913&&0.030&-0.030&&0.024&0.038&&0.741&0.888\\
&GC-VS&0.003&-0.003&&0.043&0.062&&0.934&0.935&&0.002&-0.002&&0.030&0.042&&0.932&0.948\\
&DR-SB&&-0.001&&&0.065&&&0.889&&&0.001&&&0.045&&&0.900\\
\hline
2&UA-RCT&-0.001&0.002&&0.049&0.070&&0.949&0.945&&0.000&0.000&&0.034&0.047&&0.953&0.956\\
&UA-pooled&0.091&-0.090&&0.026&0.056&&0.075&0.640&&0.091&-0.091&&0.019&0.038&&0.003&0.352\\
&GC-RCT&-0.002&0.002&&0.048&0.066&&0.944&0.941&&0.000&0.000&&0.033&0.045&&0.951&0.956\\
&GC-NI&0.027&-0.027&&0.033&0.056&&0.859&0.917&&0.027&-0.027&&0.024&0.038&&0.779&0.906\\
&GC-VS&0.002&-0.002&&0.044&0.063&&0.936&0.931&&0.002&-0.002&&0.030&0.043&&0.942&0.950\\
&DR-SB&&-0.001&&&0.065&&&0.887&&&0.002&&&0.045&&&0.908\\
\hline
3&UA-RCT&-0.001&0.002&&0.049&0.070&&0.948&0.945&&0.000&0.000&&0.034&0.047&&0.953&0.955\\
&UA-pooled&0.078&-0.077&&0.027&0.057&&0.182&0.717&&0.078&-0.078&&0.019&0.038&&0.015&0.488\\
&GC-RCT&-0.002&0.002&&0.048&0.066&&0.944&0.941&&0.000&0.000&&0.033&0.045&&0.950&0.956\\
&GC-NI&0.023&-0.023&&0.033&0.057&&0.887&0.920&&0.023&-0.023&&0.023&0.038&&0.839&0.927\\
&GC-VS&0.001&-0.001&&0.044&0.063&&0.937&0.931&&0.001&-0.001&&0.031&0.043&&0.943&0.949\\
&DR-SB&&0.000&&&0.065&&&0.894&&&0.002&&&0.045&&&0.909\\
\hline
4&UA-RCT&-0.001&0.002&&0.049&0.070&&0.948&0.945&&0.000&0.000&&0.034&0.047&&0.953&0.955\\
&UA-pooled&0.147&-0.146&&0.025&0.056&&0.000&0.247&&0.147&-0.147&&0.017&0.037&&0.000&0.031\\
&GC-RCT&-0.002&0.002&&0.048&0.066&&0.944&0.941&&0.000&0.000&&0.033&0.045&&0.950&0.956\\
&GC-NI&0.086&-0.086&&0.033&0.056&&0.259&0.649&&0.087&-0.087&&0.023&0.038&&0.035&0.395\\
&GC-VS&0.007&-0.007&&0.051&0.068&&0.907&0.932&&0.003&-0.003&&0.034&0.046&&0.935&0.949\\
&DR-SB&&-0.002&&&0.065&&&0.888&&&0.001&&&0.045&&&0.887\\
\hline
\end{tabular}
\end{center}
}
\end{table}
\end{landscape}

\renewcommand{\baselinestretch}{1.0}
\begin{table}[htbp]
\caption{Analysis of HIV example data with $\bmX=(\text{age},\text{race},\sqrt{\text{CD4}})'$: point estimates (standard errors) of $(\mu_1,\mu_0,\delta)$ from six different estimation methods (see Section \ref{sec:app} for details).}\label{hiv.rst.1}
\newcolumntype{d}{D{.}{.}{2}}
\begin{center}
\begin{tabular}{llll}
\hline
\hline
Method&\multicolumn{3}{c}{Pt.~Est.~(Std.~Err.)}\\
\cline{2-4}
&\multicolumn{1}{c}{$\mu_1$ (\%)}&\multicolumn{1}{c}{$\mu_0 (\%)$}&\multicolumn{1}{c}{$\delta$ (\%)}\\
\hline
UA-RCT&4.5 (2.2)&7.4 (2.7)&~-3.0 (3.5)\\
UA-pooled&4.5 (2.2)&8.6 (1.3)&~-4.1 (2.5)\\
GC-RCT&6.3 (2.0)&6.7 (2.6)&~-0.4 (3.0)\\
GC-NI&6.3 (2.0)&9.3 (1.5)&~-3.0 (2.3)\\
GC-VS&6.3 (2.0)&9.3 (1.5)&~-3.0 (2.3)\\
DR-SB&&&~-1.2 (2.4)\\
\hline
\end{tabular}
\end{center}
\end{table}

\renewcommand{\baselinestretch}{1.0}
\begin{table}[htbp]
\caption{Analysis of HIV example data with $X=\sqrt{\text{CD4}}$: point estimates (standard errors) of $(\mu_1,\mu_0,\delta)$ from three different estimation methods (see Section \ref{sec:app} for details).}\label{hiv.rst.2}
\newcolumntype{d}{D{.}{.}{2}}
\begin{center}
\begin{tabular}{llll}
\hline
\hline
Method&\multicolumn{3}{c}{Pt.~Est.~(Std.~Err.)}\\
\cline{2-4}
&\multicolumn{1}{c}{$\mu_1$ (\%)}&\multicolumn{1}{c}{$\mu_0 (\%)$}&\multicolumn{1}{c}{$\delta$ (\%)}\\
\hline
GC-RCT&6.8 (2.0)&6.5 (2.6)&0.3 (2.9)\\
GC-NI&6.8 (2.0)&10.0 (1.5)&~-3.2 (2.2)\\
GC-VS&6.8 (2.0)&9.5 (2.5)&~-2.6 (2.9)\\
\hline
\end{tabular}
\end{center}
\end{table}

\end{document}